\documentclass[pra,aps,showpacs,showkeys,groupedaddress,superscriptaddress,twocolumn]{revtex4-1}

\usepackage[utf8x]{inputenc}
\usepackage{color}
\usepackage{bbm} 
\usepackage{amsfonts,amsmath,amssymb,stmaryrd}
\usepackage{graphicx}
\usepackage{subfigure}  % use for side-by-side figures
\usepackage{hyperref}
\usepackage{epsfig}
\usepackage{mathrsfs}
\usepackage{verbatim}
\usepackage{centernot}

\newcommand{\ket}[1]{|#1\rangle}

\usepackage{array}

\usepackage{bm}	% bold in math mode

\begin{document}
%%%%%%%%%%%%%%%%%%%%%%%%%%%%%%%%%%%%%%%%%%%%%%%%%
\title{Quantized single-particle Thouless pump 
induced by topology transfer from a 
Chern insulator at finite temperature}
%%%%%%%%%%%%%%%%%%%%%%%%%%%%%%%%%%%%%%%%%%%%%%%%%

\author{Lukas Wawer}
\affiliation{Department of Physics and Research Center OPTIMAS, University of Kaiserslautern, 67663 Kaiserslautern, Germany}
\author{Razmik Unanyan}
\affiliation{Department of Physics and Research Center OPTIMAS, University of Kaiserslautern, 67663 Kaiserslautern, Germany}
\author{Michael Fleischhauer}
\affiliation{Department of Physics and Research Center OPTIMAS, University of Kaiserslautern, 67663 Kaiserslautern, Germany}
%\author{Michael Fleischhauer}
%\affiliation{Department of Physics and Research Center OPTIMAS, University of Kaiserslautern, 67663 Kaiserslautern, Germany}

%%%%%%%%%%%%%%%%%%%%%%%%%%%%%%%%%%%%%%%%%%%%%%%%%

\begin{abstract}
Quantized particle or spin transport upon cyclic parameter variations, determined by topological invariants, is a key signature of 
Chern insulators in the ground state.
While measurable many-body observables exist that preserve the integrity of topological invariants also at finite temperature, quantized transport is generically lost.
% restricted to low temperatures. 
We here show that a coupling of a one-dimensional 
Chern insulator at arbitrary finite temperature to an auxiliary lattice can induce quantized transport determined by the finite-temperature invariant. We show for the example of a Rice-Mele model that the spatial distribution of a single particle in the auxiliary chain moves by a quantized number of unit cells in a Thouless cycle when subtracting a spatially homogeneous offset even at a temperature exceeding the band gap.
\end{abstract}
\pacs{}

\date{\today}
\maketitle

%%%%%%%%%%%%%%%%%%%%%%%%%%%%%%%%%%%%%%%%%%%%%%%%%

%%%%%%%%%%%%%%%%%%%%%%%%%%%
%%%%%%%%%%%%%%%%%%%%%%%%%%%
\section{Introduction}
%%%%%%%%%%%%%%%%%%%%%%%%%%%
%%%%%%%%%%%%%%%%%%%%%%%%%%%

Topology has become an important concept to classify ground states of many-body quantum systems \cite{Klitzing-PRL-1980,Kosterlitz-JPC-1973,Laughlin-PRL-1983,Arovas-PRL-1984,Xiao-RMP-2010,Hazan-Kane-RMP-2010,Wen-RMP-2017,Haldane-PRL-1988,Wang-PRL-2013,TKNN-PRL-1982,Thouless-PRB-1983,Niu-JPhysA-1984,Niu-PRB-1985,Oshikawa-PRL-2000,King-Smith-PRB-1993,Tao-PRB-1984,Zheng-PRB-2016,ExtSLBHM,Kane-Mele_PRL-2005,Fu-Kane-PRB-2006,Sheng-PRL-2006,Qi-Wu-Zhang-PRB-2006,Altland-PRB-1997,Schnyder-PRB-2008,Ryu-NJPhys-2010,Li-PRB-2017}. A gapped band structure is called topological if its Bloch eigenstates show, colloquially speaking, non-trivial "twists" as a function of system parameters. This global property, characterized by integer invariants, has a high degree of robustness to perturbations or deformations of the Hamiltonian. The existence of topological invariants is also the origin of a number of practically important features of these systems. An example is the strictly quantized particle or spin transport in insulating states upon cyclic parameter variations, which manifests itself in the one-dimensional Thouless pump \cite{TKNN-PRL-1982,Niu-JPhysA-1984,Niu-PRB-1985,Thouless-PRB-1983,Wang-PRL-2013,Li-PRB-2017} or in the quantization of conductivities in the Quantum-Hall \cite{Klitzing-PRL-1980,Kosterlitz-JPC-1973,Arovas-PRL-1984,Laughlin-PRL-1983,Haldane-PRL-1988,TKNN-PRL-1982,Niu-PRB-1985} and Quantum-Spin-Hall effects \cite{Kane-Mele_PRL-2005,Fu-Kane-PRB-2006,Sheng-PRL-2006,Qi-Wu-Zhang-PRB-2006}. The  quantization of these observables is however restricted to ground or low-temperature states of the many-body system and the mixedness of a quantum state is seen as a general adversary to topological quantization. 

Extending topology to finite-temperature or non-equilibrium states is a long standing quest \cite{Niu-JPhysA-1984,Bardyn-NJP-2013,Linzner-PRB-2016,Bardyn-2017,Grusdt-PRB-2017,Bardyn-PRX-2018,Wawer-PRB-2021,Wawer-PRA-2021,Wawer-arxiv-2021,Mink-PRB-2019,Viyuela-PRL-2014,Viyuela-PRL-2014b,Huang-PRL-2014,Uhlmann-Rep-Math-Phys-1986,Budich-Diehl-PRB-2015,Nieuwenburg-PRB-2014,Altland-PRX-2021,Lieu-PRL-2020}. In \cite{Bardyn-NJP-2013} a topological classification of mixed states of non-interacting fermions, which are Gaussian, was suggested in terms of the ground state of the so-called fictitious Hamiltonian, fully characterizing the Gaussian state. More recently  many-body correlators were identified that generalize topological invariants to mixed states of systems with broken time-reversal (TR) symmetry \cite{Linzner-PRB-2016,Bardyn-PRX-2018,Wawer-PRB-2021} as well as with TR invariance \cite{Wawer-arxiv-2021}, and which support this classification. While these many-body correlators can be measured and corresponding detection schemes have been proposed \cite{Bardyn-PRX-2018}, it remains an open question if topological quantization of observables with more direct practical relevance survives for finite-temperature or non-equilibrium systems. In the present paper we address this question for a special class of $1+1$ dimensional, Chern  insulators of non-interacting fermions in a thermal state with temperatures below and above the band gap.

In a recent paper \cite{Wawer-PRA-2021} we have shown that a one-dimensional lattice system weakly coupled to an auxiliary, commensurate lattice of non-interacting fermions can transfer its topological properties to the auxiliary system at zero temperature. As a consequence of this "topology-transfer" a quantized transport could be observed in the auxiliary system upon an adiabatic cyclic variation of system parameters of the original system associated with its topological winding or Chern number. 
We here show for the simplest example of a $1+1$-dimensional model with broken TR symmetry, the Rice-Mele model (RMM) \cite{Rice-Mele}, that the topology transfer scheme also leads to a quantized transport of a single particle in the auxiliary chain if the RMM is in a finite-temperature state, provided an appropriate initial state of the auxiliary particle is prepared.

%%%%%%%%%%%%%%%%%%%%%%%%%%%
%%%%%%%%%%%%%%%%%%%%%%%%%%%
\section{Topology transfer}
%%%%%%%%%%%%%%%%%%%%%%%%%%%
%%%%%%%%%%%%%%%%%%%%%%%%%%%

\begin{figure}[htb]
	\begin{center}
	\includegraphics[width=0.7\columnwidth]{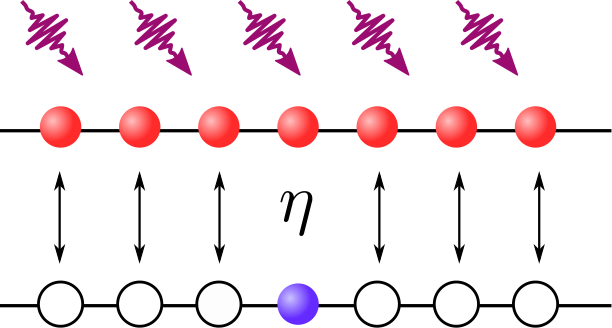}
	\end{center}
	\caption{Topology transfer scheme: The original  chain of fermions (top) is weakly coupled to an auxiliary chain of otherwise non-interacting  fermions (bottom). In the absence of the coupling the auxiliary fermions are deeply bound with negligible hopping between lattice sites. The coupling is diagonal in momentum space and conserves the particle numbers in both chains. We here consider a single, initially well localized particle in the auxiliary chain.}
	\label{fig:transfer}
\end{figure}

Let us consider a one-dimensional lattice of non-interacting fermions with particle number conservation consisting of $L$ unit cells and with periodic boundary conditions at some finite temperature $T$. The lattice constant is $a=1$ and we set $\hbar =1$ throughout this work. 
The operators $\hat c_{\mu,j},\hat c_{\mu,j}^\dagger$
describe the annihilation and creation of a fermion in the $j$th unit cell and the index $\mu\in \{1,\dots,p\}$ denotes a possible internal degree of freedom. Assuming translational invariance for simplicity, the Hamiltonian can be written in momentum space as
\begin{equation}
H_{s}= \sum_k \sum_{\mu,\nu=1}^p  \hat c_\mu^\dagger(k) \, {\sf h}_{\mu\nu}(k) \,  \hat c_\nu(k).\label{eq:H}
\end{equation}
with $k$ being the lattice momentum. ${\sf h}(k)$ is the single-particle Hamiltonian in Bloch space which we assume to have a non-trivial topological band structure. This system is now weakly coupled to a commensurate lattice of otherwise non-interacting auxiliary fermions with respective annihilation and creation operators $\hat a_\mu(k)$ and $\hat a^\dagger_\mu(k)$ 
as indicated in Fig.\ref{fig:transfer},
\begin{eqnarray}
H &=& H_{s} + H_\eta ,\label{eq:H_tot}\\
H_\eta &=& \eta \sum_k \sum_{\mu,\nu=1}^p   \hat  c_\mu^\dagger(k) \hat c_\nu(k) \hat a_\mu^\dagger(k) \hat a_\nu(k),\label{eq:H_eta}
\end{eqnarray}
where we have assumed a unit cell of $p$ sites.
Obviously the number of fermions in both chains is individually conserved.

%%%%%%%%%Fig. coupling scheme 1D%%%%%%%%%%%%%%%%%%
%%%%%%%%%%%%%%%%%%%%%%%%%%%

In Ref.\cite{Wawer-PRA-2021} we have shown that 
an adiabatic cyclic variation of parameters of $H_s(t)$,
known as a Thouless pump, induces a strictly quantized particle transport in the auxiliary chain if the system is in the ground state. 
In the present paper we are going to show that this transfer remains intact if the topological lattice is in a finite-temperature state. 

Let us first discuss the effect of the auxiliary chain to the topological lattice. We note that in an insulating state of the isolated Chern insulator, particle transport is entirely due to non-adiabatic terms and the action of the auxiliary chain has to be compared to those terms.
In the instantaneous eigenbasis of ${\sf h}(k,t)$ the time evolution of the Chern insulator is described by the effective Hamiltonian
\begin{equation}
    {\sf h}_\mathrm{eff}(k) = \mathrm{diag}\bigl(\varepsilon_\mu(k)\bigr)
+ \partial_t     
    {\sf r}(k) \cdot {\sf r}^{-1}(k) 
\end{equation}
where ${\sf r}(k,t)$ is the matrix that diagonalizes ${\sf h}(k,t)$. The second term describes non-adiabatic contributions and scales as $\tau^{-1}$ where $\tau$ is the cycle time. In order for a Thouless pump in the combined system to be adiabatic $\tau$ has to be chosen large enough, such that $\eta \tau \gg 1$, where $\eta$ is strength of the weak coupling to the auxiliary chain. On the other hand the perturbation created by the auxiliary chain to the dynamics of $H_s$ is on the order of $\eta \overline{n}_k$ where $\overline{n}_k$ is the typical number of auxiliary particles per lattice momentum $k$. 
Thus with respect to the adiabatic transport in the original lattice the coupling to the auxiliary chain is a non-negligible perturbation if $\overline{n}_k\sim {\cal O}(1)$.

%%%%%%%%%Fig. RMM transport%%%%%%%%%%%%%%%%%%
%\begin{figure}[htb]
%	\begin{center}
%	\includegraphics[width=1.0\columnwidth]{Fig_2.png}
%	\end{center}
%	\caption{Particle transport during a Thouless pump cycle of duration $\tau$ in a RMM coupled to a fermion chain initially prepared in the ground state and with half filling in both chains. Shown in a) is the transport in the auxiliary chain, and in b) in the RMM.}
%	\label{fig:RMM-transport}
% \end{figure}
%%%%%%%%%%%%%%%%%%%%%%%%%%%

% This back-coupling is illustrated in Fig. \ref{fig:RMM-transport} where we have plotted the transported number of particles for a RMM coupled to an auxiliary fermion chain. 

This back-action effect can be eliminated either by
using several identical topological lattices coupled to a single auxiliary chain or by considering a single, initially well localized auxiliary particle. In the latter case, which we shall discuss in the following, the characteristic particle number per mode scales inversely with system size and $H_\eta$ becomes an irrelevant perturbation to the adiabatic particle transport in the Chern insulator.

%%%%%%%%%%%%%%%%%%%%%%%%%%%%%%%%%%%%%
\section{Single-particle transport in topological band structure}\label{sec:single-particle}
%%%%%%%%%%%%%%%%%%%%%%%%%%%%%%%%%%

Let us first discuss a single particle in an given one-dimensional topological band structure with lattice constant $a=1$ and length $L$ with periodic boundary conditions. The system is in general described by the single-particle Hamiltonian matrix 
%
%\begin{equation}
%H=\sum_{k=-L/2}^{L/2} \vert k\rangle \langle k\vert \otimes %{\sf h}(k),\label{hamiltonian}
%\end{equation}
%
% where 
${\sf h}(k)$ 
% is the Hamiltonian matrix 
in momentum space, describing the internal dynamics of a single unit cell.  
% $\left\vert k\right\rangle $ are plain waves with Wannier projections $\langle n \vert k\rangle=\frac{1}{\sqrt{L}}\exp\left(\frac{2\pi i}{L}kn\right)$, $\langle k^{\prime} \vert k\rangle =\delta_{k,k^{\prime}}$. Here $k$ is the quasi-momentum of the particle and $n$ is the coordinate of the unit cell on the lattice. 
The solution of the single-particle Schrödinger equation
%
%\begin{equation}
%    i\frac{\partial}{\partial t}\left\vert \Phi\left( %  t\right)  \right\rangle =H\left\vert \Phi\left(   % t\right)  \right\rangle ,\label{schrodinger}%
% \end{equation}
%
with initial condition $\left\vert \Phi\left(  t=0\right)  \right\rangle
=\left\vert \Phi_{0}\right\rangle $, can be represented as
\begin{equation}
    \left\vert \Phi\left(  t\right)  \right\rangle =\frac{1}{\sqrt{L}} \sum_{k=-L/2}^{L/2} C_{k}
    \left\vert k\right\rangle \otimes
    \left\vert \phi(k,t)  \right\rangle, 
\end{equation}
where $\left\vert C_{k}\right\vert ^{2}$ gives the
initial quasi-momentum probability distribution. 
%The norm of the wave function is constant in time, which implies $\frac{1}{L}\sum_{k=-L/2}^{L/2} \left\vert C_{k}\right\vert ^{2}=1.$ 
Since lattice momentum is conserved, the time evolution factorizes and is governed by
\begin{equation}
i\frac{\partial}{\partial t}\left\vert \phi(k,t)  \right\rangle ={\sf h}(k)\,  \left\vert \phi(k,t)  \right\rangle. \label{band_Hamiltonian}
\end{equation}
Now lets us consider a slow and cyclic time variation ${\sf h}(k)\to {\sf h}(k,t)$ such that the single particle band gap is not closed, and  ${\sf h}(k,t)  ={\sf h}(k,t+\tau)$.
We are interested in the particle transport over a single  period $\tau$. 
%The momentum depends of ${\sf h}(k,t)$ causes tunneling transitions between different superlattice sites. 
In the following, we will only analyze the case when the initial state $\left\vert \phi(k,0)\right\rangle$ coincides with one of the adiabatic eigenstates of ${\sf h}(k,0)$ and corresponds to the same Bloch band (e.g. the ground state) for all values of $k$.

The probability to find a particle in the unit cell $n$, without specification of internal states, is
\begin{equation}
    P_{n}\left(t\right)  =\left\langle u_n(t)  \right. \left\vert u_n(t)\right\rangle,
\end{equation}
where
\begin{eqnarray}
    \left\vert u_n(t) \right\rangle &=& \langle n\vert \Phi(t)\rangle \nonumber\\
    &=&\frac{1}{\sqrt{L}}\sum_{k=-L/2}^{L/2} C_{k}\left\langle n\right.  \left\vert k\right\rangle \left\vert \phi(k,t)  \right\rangle \\
    &=&\frac{1}{L}\sum_{k=-L/2}^{L/2} C_{k}
    \exp\left\{\frac{2\pi i}{L} kn\right\}\left\vert \phi(k,t)  \right\rangle .\nonumber
\end{eqnarray}
Further we will examine the case of large system size $L\rightarrow\infty$. In this limit, the sum can be replaced by an integral, $q= 2\pi k/L$,
\begin{equation}
    \left\vert u_{n}(t)  \right\rangle =\frac{1}{2\pi} \int_{-\pi}^{\pi}\!\! dq\,\, C(q) e^{inq}  \left\vert \phi(q,t) \right\rangle .\label{Wannier}%
\end{equation}
The shift of the center of mass (COM) of the particle upon an adiabatic cyclic change of Hamiltonian parameters without closing the single particle band gap, $R(t)=\langle \hat{x} \rangle_0 - \langle \hat{x} \rangle_t$,  can be written as
\begin{align}
R(t)  &=\sum_{n=-\infty}^{\infty} n\,  P_n(t)=\int_{-\pi}^\pi\!\! dq \int_0^\tau\!\! dt\,\vert C(q)\vert^2 \frac{\partial \epsilon(q,t)}{\partial q}\nonumber\\
&+
\frac{1}{2\pi} \int_0^{t} \!\! \mathrm{d}t^\prime\, \int_{-\pi}^{\pi} \!\! \mathrm{d}q \, \, \vert C(q) \vert^2 \mathcal{A}(q,t^\prime), \label{Polarization}%
\end{align}
where for simplicity we assumed the initial condition
$\langle \hat{x} \rangle_0=0$. The first term describes the dynamical contribution in the adiabatically varying band structure in the chosen Bloch band of energy $\epsilon(q,t)$. The second term describes geometric contributions with $\mathcal{A}(q,t) = i\left\langle \phi(q,t)\right.\left\vert\partial_q \phi(q,t)  \right\rangle$ being the Berry connection. If we choose $\vert C(q)\vert^2$ equal for all lattice momenta $q$, the dynamical term vanishes and the term $\vert C(q)\vert^2$ can be pulled out of the second integral, which is then just the topological winding number (Chern number) of the band structure. Thus after one cycle $\tau$ the shift of the center of mass $R(\tau)$ is quantized. 
In the following we assume $C(q) =e^{-in_{0}q}$ ($n_{0}$ denoting the initial unit cell). In this case the state vector $\left\vert u_{n}(t=0)\right\rangle $ coincides with the Wannier function of unit cell $n_0$ in the given Bloch band (e.g. the lowest band), which we set to $n_0=0$. 
Analogously, we may calculate the uncertainty $\Delta R^2(t)= \langle \Delta \hat{x}^2 \rangle_t = \langle \hat{x}^2\rangle_t - \langle \hat{x}\rangle_t^2$ of the center of mass coordinate. After a simple calculation we eventually obtain the following form
\begin{align}
    \Delta R^{2}(t)  &=\sum_{n=-\infty}^{\infty} n^{2}P_n(t) - R(t)^2 \\
    &=\frac{1}{2\pi}\int_0^t\!\!\mathrm{d}t^\prime\int _{-\pi}^{\pi} \!\! \mathrm{d}q\, \left\langle \partial_q\phi(q,t^\prime)\right. \left\vert \partial_q\phi(q,t^\prime)\right\rangle- R(t)^2. \label{Uncertainty00}
\end{align}
After one cycle   $\Delta R^{2}\left(  \tau\right) =A+B  $, where
\begin{equation}
    A =\frac{1}{2\pi}\int_0^\tau\!\! \mathrm{d}t^\prime\int_{-\pi}^{\pi}\!\!  \mathrm{d}q\, \left\langle \partial_q\phi(q,t^\prime)\right\vert \Pi(q,t^\prime) \left\vert \partial_q\phi(q,t^\prime) \right\rangle \geq0\label{estimation_B}%
\end{equation}
and
\begin{equation}
    B =\frac{1}{2\pi}\int_0^\tau\!\! \mathrm{d}t^\prime\int_{-\pi}^{\pi}\!\!\mathrm{d}q \, \left\vert \left\langle \phi\left(q,t^\prime\right)  \right.  \left\vert \partial_q\phi(q,t^\prime)\right\rangle \right\vert^{2} \, -R(\tau)^2 \geq0. \label{estimation_A}%
\end{equation}
Here $\Pi(q,t)  =\mathbbm{1}-\left\vert \phi(q,t) \right\rangle \left\langle \phi(q,t) \right\vert $ is the projection operator onto the orthogonal space to $\left\vert \phi(q,t)  \right\rangle $. In the adiabatic limit $A$ does not vanish and has the geometric interpretation of "band flatness".
The second term, $B$, includes a geometric dispersion, which does not depend on the period $\tau$. 

%
%%%%%% COM and variation of RMM %%%%% 
\begin{figure}[htb]
	\begin{center}
	\includegraphics[width=0.9\columnwidth]{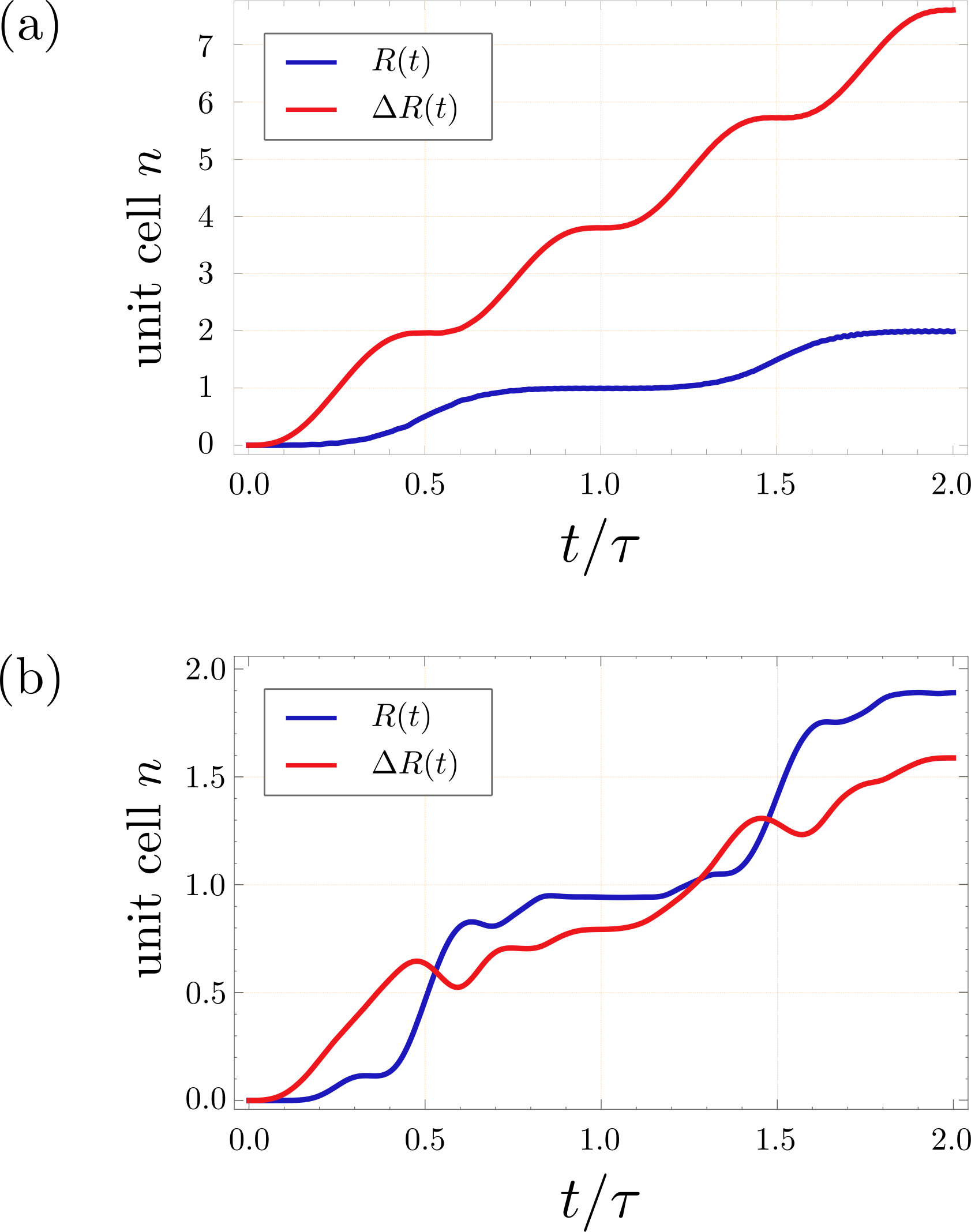}
	\end{center}
	\caption{Center-of-mass $R(t)$ and dispersion $\Delta R\left(  t\right)  $ for a RMM with hopping amplitudes $t_{1,2}=-\frac{\Omega_{0}}{4}\left(1\pm\cos\left(\frac{2\pi t}{\tau}\right)\right)  $ and offset $\Delta=\frac{\Omega_{0}}{2}\sin\left( \frac{2\pi t}{\tau}\right)$ for an adiabatic Thouless pump (a) $\Omega_0 \tau =100$ and (b) $\Omega_0 \tau =20$.}
	\label{fig:RMM-single-particle}
\end{figure}
%%%%%%%%%%%%%%%%%%%%%%%%%%%%%%%
%
In Fig.\ref{fig:RMM-single-particle} we have illustrated the time evolution of the center-of-mass $R(t)$ as well as the dispersion $\Delta R(t)$ for a Rice Mele model \cite{Rice-Mele}. This model is used also in the remainder of the paper to illustrate our findings. It has a unit cell  of two sites labelled $(\mathrm{A})$ and $(\mathrm{B})$, and has alternating hopping amplitudes $t_1$ and $t_2$ as well as energy offsets $\pm\Delta$
\begin{align}
	& \quad H_\mathrm{RM} =\label{eq:RMM} \\
	%-\sum_k (t_1 + t_2 \mathrm{e}^{-ik}) \hat{c}_{\mathrm{A}}^\dagger(k) \hat{c}_{\mathrm{B}}(k) + \mathrm{h.c.} + \Delta \sum_k \left( \hat{c}_{\mathrm{A}}^\dagger(k)\hat{c}_{\mathrm{A}}(k) - \hat{c}_{\mathrm{B}}^\dagger(k) \hat{c}_{\mathrm{B}}(k) \right) \nonumber \\
		&= \sum_k \begin{pmatrix}
		\hat{c}_{\mathrm{A}}^\dagger(k) \\
		\hat{c}_{\mathrm{B}}^\dagger(k)
		\end{pmatrix}^\mathrm{T}
		\begin{pmatrix}
		\Delta & -t_1-t_2 \mathrm{e}^{-ik} \\ 
		 -t_1-t_2 \mathrm{e}^{ik} & -\Delta
		\end{pmatrix}
		\begin{pmatrix}
		\hat{c}_{\mathrm{A}}(k)\\
		\hat{c}_{\mathrm{B}}(k)
		\end{pmatrix}.\nonumber
	\end{align}
Since the dynamical dispersion of the wave packet increases with the length of the cycle period $\tau$, the adiabatic limit required for the quantization of the COM motion is usually associated with a large spread of the wave function. This can be seen from Fig.\ref{fig:RMM-single-particle}. In the adiabatic limit, Fig.\ref{fig:RMM-single-particle}a, there is
quantization of the COM shift but the spread is large, while going away from that limit,  Fig.\ref{fig:RMM-single-particle}b, the spread is reduced but quantization is lost.
By flattening the energy surface of the adiabatic transfer state one can reduce the dynamical dispersion of the particle transport and only the geometric dispersion remains.
%\textcolor{red}{ I think we have to explain this better by providing an appendix describing the final expressions for B and A terms in the adiabatic approximation. This appendix is ready. However, I am not sure about that. On the other hand I am afraid that a more detailed discussion about variance will distract readers from the main idea of the article}

%In other words the adiabatic transport of a single particle on one-dimensional superlattice is not quantized but the transport of the center of mass $\langle \hat{x} \rangle$

%This we can see in Fig. \ref{fig:transfer-1particle_RM}, where we calculated the particle distribution of a single particle in a RMM after two Thouless pump cycles. One sees that for low adiabatic condition $\Omega_0\tau=6$ the peak of the distribution is sharp ($\Delta R^2(\tau)=0.54$, however the shift of the center of mass is not quantized $R(\tau)=0.93$. In the adiabatic case $\Omega_0\tau=40$ the center of mass $R(\tau)=0.99$ is quantized. Due to the "unflatness" of the RMM energy bands the term $B(\tau)$ is large and we see a strong spreading of the probability distribution.

%%%%%%%%%%%%%%%%%%%%%%%%%%%%%%%%%
%%%%%%%%%%%%%%%%%%%%%%%%%%%%%%%%%
\section{Topology transfer to a single particle}
%%%%%%%%%%%%%%%%%%%%%%%%%%%%%%%%%
%%%%%%%%%%%%%%%%%%%%%%%%%%%%%%%%%

In order to analyze the effect of topology transfer from a Chern insulator at finite temperature to a single particle, it is instructive to decompose the total Hamiltonian \eqref{eq:H_tot} in the form
\begin{equation}
H=H_0 + H_1 \label{eq:H-alt}
\end{equation}
where
\begin{equation}
H_0 = H_{s} + \eta \sum_k \sum_{\mu,\nu=1}^p   \bigl\langle \hat  c_\mu^\dagger(k) \hat c_\nu(k)\bigr\rangle  \hat a_\mu^\dagger(k) \hat a_\nu(k)\label{eq:H0}
\end{equation}
contains the system Hamiltonian $H_s$ and the mean-field interaction Hamiltonian, where ${\sf m}_{\mu\nu}= \langle \hat  c_\mu^\dagger(k) \hat c_\nu(k)\rangle$ is evaluated in the initial thermal state of $H_{s}$. The second term in \eqref{eq:H-alt} formally describes the coupling of the auxiliary chain to fluctuations in the original system
\begin{equation}
H_1 =  \eta \sum_k \sum_{\mu,\nu=1}^p   \Bigl(\hat  c_\mu^\dagger(k) \hat c_\nu(k) - \bigl\langle \hat  c_\mu^\dagger(k) \hat c_\nu(k)\bigr\rangle\Bigr)
\hat a_\mu^\dagger(k) \hat a_\nu(k)\label{eq:H1}
\end{equation}
and is responsible for the buildup of entanglement between the two chains.

Eq.\eqref{eq:H0} describes the evolution of auxiliary fermions under an effective single-particle Hamiltonian, which corresponds to the fictitious Hamiltonian of the original topological model in the Gaussian thermal state
\begin{equation}
\rho = \frac{1}{Z} \exp\left\{-\beta\sum_k \hat{\mathbf{c}}^\dagger(k)\, ({\sf h}(k)-\mu)\, \hat{\mathbf{c}}(k)\right\}\label{eq:Gauss}
\end{equation}
where $\beta = 1/(k_B T)$ and $\mu$ is the chemical potential. Since this state is Gaussian, correlations can easily be calculated leading to an effective single-particle Hamiltonian proportional to
\begin{equation}
\langle c_\mu^\dagger(k) c_\nu(k)\rangle= \frac{1}{2}\left[ 1- \tanh\left(\frac{\beta({\sf h}(k)-\mu)}{2}\right)\right]_{\nu,\mu},\label{eq:m}
\end{equation}
which has the same eigenstates than ${\sf h}(k)$. Moreover the  energy bands are flattened as compared to those of ${\sf h}(k)$. In fact in the limit $T\to 0$ the
band dispersion vanishes completely.

%%%%%%%%%%%%%%%COM + dispersion mean-field%%%%%%%
\begin{figure}[htb]
	\begin{center}
	\includegraphics[width=0.9\columnwidth]{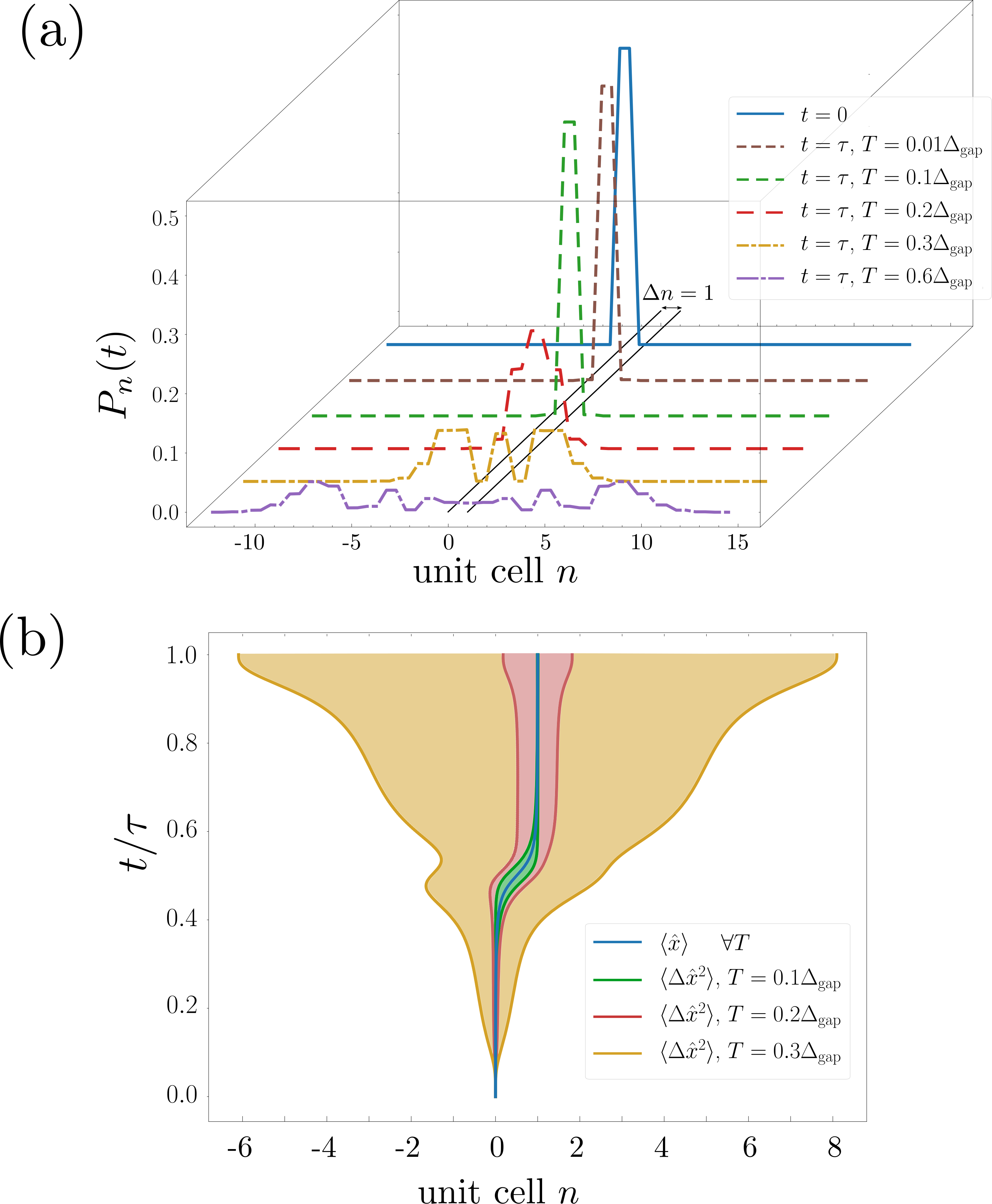}
	\end{center}
	\caption{(a) Particle distribution $P_n(t)$ of auxiliary particle after one Thouless pump cycle $\tau=100$ for a finite-temperature RMM in the mean-field limit with coupling $\eta=0.01 \Delta_\mathrm{gap}$. The mean-field Hamiltonian is an effective RM Hamiltonian with hoppings $\tilde{t}_{1,2}(k) =\frac{\eta}{2\varepsilon(k)}		\tanh\left(\frac{\beta \varepsilon(k)}{2}\right)\, t_{1,2}$ and staggered potential $\tilde{\Delta}(k)=\frac{\eta}{2\varepsilon(k)}\tanh\left(\frac{\beta \varepsilon(k)}{2}\right)		\, \Delta$, where $\varepsilon(k)=\sqrt{\Delta^2 +t_1^2+t_2^2+2t_1t_2\cos(2\pi k/L)}$ is the energy and $t_{1,2}=(1\pm\cos(2\pi t/\tau))$, $\Delta=-2\sin(2\pi t/\tau)$ are the parameters of the RMM. (b) Center of mass $R(t)=\langle \hat{x} \rangle_t$ and spread of the wave packet $\Delta R^2(t) = \langle \Delta \hat{x}^2 \rangle_t$. While the motion of the center of mass is strictly quantized for all temperatures the spreading increases with increasing temperature.}
	\label{fig:transfer-1particle_MF}
\end{figure}
%%%%%%%%%%%%%%%%%%%%%%%%%%%%%%%

%%%%%%%%%%%%%%%%%%%%%%%%%%%%%%%%%%%%%
\subsection{Topology transfer in mean-field approximation}
%%%%%%%%%%%%%%%%%%%%%%%%%%%%%%%%%%

The discussion of the Sect.\ref{sec:single-particle} can straightforwardly be applied to the topology transfer scheme in mean-field approximation. In this limit, described by $H_0$ alone, the auxiliary particle evolves under the fictitious Hamiltonian of the finite-temperature Chern insulator. As seen from eq.\eqref{eq:m}, the eigenstates of the fictitious Hamiltonian show a topological winding at any finite temperature with a winding- or Chern-number given by that of the ground state of the topological model. Thus we expect a strictly quantized transport of the center of mass $R=\langle \hat x\rangle$ of the auxiliary particle in a single Thouless cycle in the adiabatic limit. We see in Fig. \ref{fig:transfer-1particle_MF} for the example of a finite-temperature RMM that the particle transport $R(\tau)$ is indeed quantized for any temperature. Furthermore for temperatures small compared to the gap of the topological model the spread of the wave packet in position space $\Delta R^2(\tau)$ is small but increasing with growing temperature, because the reduced flatness of the mean-field energy bands with temperature.
%
%\begin{eqnarray}
%    \varepsilon(k)_n^\textrm{fict} = \frac{\eta}{2}\left[1-\tanh\left(\frac{\beta(\varepsilon_n(k)-\mu)}{2}\right)\right].
%\end{eqnarray}
%

%%%%%%%%%%%%%%%%%%%%%%%%%%%
%%%%%%%%%%%%%%%%%%%%%%%%%%%
\subsection{Full dynamics}
%%%%%%%%%%%%%%%%%%%%%%%%%%%
%%%%%%%%%%%%%%%%%%%%%%%%%%%

\begin{figure}[htb]
	\begin{center}
	\includegraphics[width=0.9\columnwidth]{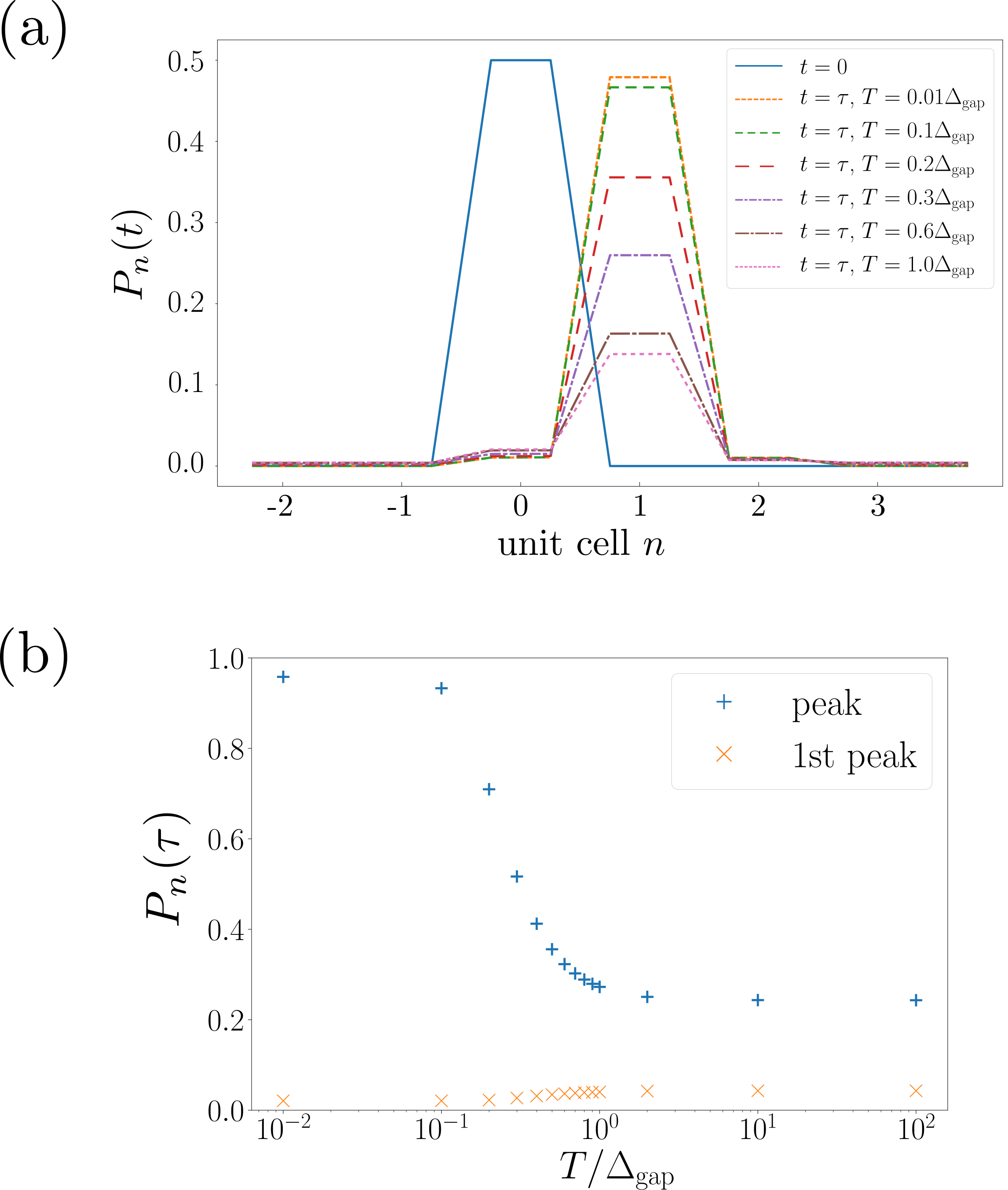}
	\end{center}
	\caption{(a) Particle distribution $P_n(t)$ after one Thouless pump cycle in the auxiliary system of the full Hamiltonian with coupling $\eta=0.01 \Delta_\mathrm{gap}$. Despite increasing temperature $T$ one observes a clear and single peak which motion is strictly quantized. (b) Particle distribution of the peak and the next nearest neighbour peaks. For all temperatures the peak is separated by its neighbouring peaks.}
	\label{fig:transfer-1particle}
\end{figure}

\begin{figure}[htb]
	\begin{center}
	\includegraphics[width=0.9\columnwidth]{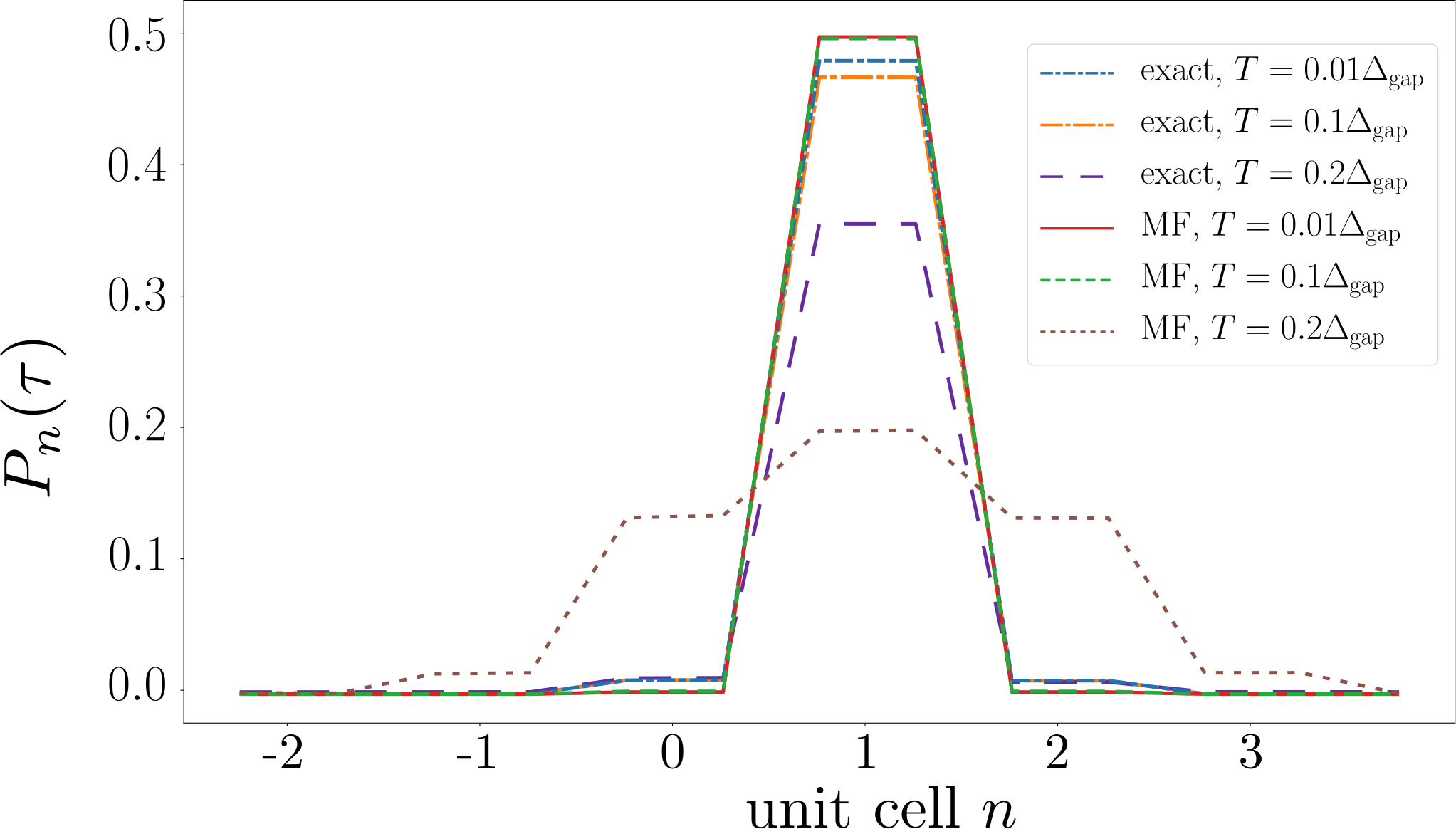}
	\end{center}
	\caption{Comparison of the particle distribution $P_n(t)$ after one Thouless pump cycle in the mean-field and the exact model with coupling $\eta=0.01 \Delta_\mathrm{gap}$.}
	\label{fig:transfer-1particle_Vergleich_MF}
\end{figure}

We now turn to the discussion of the full problem, i.e. including the fluctuation coupling, $H_1$. We again consider a RMM at finite temperature coupled to a single fermion according to eq.\eqref{eq:H}. Initially the auxiliary particle is prepared in a single unit cell in the lowest Wannier state. We numerically calculate the time evolution of the probability distribution $P_n(t)$ to find the auxiliary particle at lattice site $n$. Fig.\ref{fig:transfer-1particle}a shows the resulting distribution before and after one Thouless cycle in the adiabatic limit for different temperatures of the RMM. At $t=0$ the Wannier state has equal weight at the two sites of the unit cell.
%
%%%%%%%%%particle distribution full system%%%%%
%%%%%%%%%%%%%%%%%%%%%%%%%%%%%%%
%
One recognizes that after one Thouless cycle the peak of the probability distribution of the auxiliary particle is shifted by \emph{exactly} one unit cell for all temperatures, even above the single particle gap. Different from the mean-field limit there is a small spreading of the distribution even at $T=0$, while the COM of the distribution shifts by exact one unit cell as in the mean field case. At finite values of $T$ the dispersion of the probability distribution differs substantially from the mean field behaviour. As can be seen in Fig.\ref{fig:transfer-1particle_Vergleich_MF} the spread is much less pronounced in the full model than in the mean-field limit. 
%
%%%%%%%%Comparison mean-field and exact system%%%%
%%%%%%%%%%%%%%%%%%%%%%%%%%%%%%%
%
However, increasing the temperature there is a small flow of probability away from the peaks into the wings of the distribution creating a homogeneous background of probability in all unit cells. This homogeneous offset causes a small shift of the center of mass of the particle distribution from exact integer values. We numerically verified, however, that when the offset is subtracted and the resulting distribution renormalized, the COM shift is again perfectly quantized. Moreover, as seen in Fig.\ref{fig:transfer-1particle}b, the probability weights in the main and second-to-main peaks saturate at a rather large value when increasing the temperature beyond the single-particle gap $\Delta_\textrm{gap}$ of the RMM. We conclude that the topology transfer from a high-temperature topological model leads to  quantized transport of the peak of the probability distribution of a single auxiliary particle. While there is a small probability flow to all other unit cells, this flow saturates and only leads to 
constant background which can be well separated.

%%%%%%%%%%%%%%%%%%%%%%%%%%%
%%%%%%%%%%%%%%%%%%%%%%%%%%%
\section{Summary and Conclusion}
%%%%%%%%%%%%%%%%%%%%%%%%%%%
%%%%%%%%%%%%%%%%%%%%%%%%%%%
In the present paper we have shown that topological properties of a one-dimensional Chern insulator at arbitrary temperature can be transferred to a single particle in a second, auxiliary chain, weakly coupled to the first. An adiabatic cyclic variation of parameters of the Chern insulator 
leads to a quantized transport of the probability distribution of the auxiliary particle. While the topologial quantization of transport in the Chern insulator itself is lost for temperatures close or above the single-particle gap, its topological properties remain
encoded in the covariance matrix of single-particle correlations.
The coupling is constructed in such a way that the auxiliary particle experiences an effective mean-field Hamiltonian, called fictitious Hamiltonian, which is given by this covariance matrix. As a consequence an adiabatic cyclic variation of parameters induces in the mean-field limit a quantized motion of the center of mass determined by the Zak-phase winding number of the fictitious Hamiltonian, which is an integer-valued topological invariant. 

We first considered the mean-field limit corresponding to a single particle in an effective topological band structure. We showed that the change of the COM of an initially well localized particle is always quantized for an adiabatic Thouless pump cycle. At the same time there is a substantial spread of the particle distribution 
which increases with the cycle time. This spread is minimized to a geometric contribution if the band structure has flat bands. At low temperatures of the Chern insulator, the effective mean-field Hamiltonian of the auxiliary particle has a flat spectrum and there
is only a small spread of the probability distribution during a Thouless cycle.

The exact adiabatic time evolution of the system differs from the mean-field behaviour at larger temperatures, which we could however only investiagte numerically for a finite-temperature RMM model coupled weakly to a single particle. Surprisingly and different from the mean field limit, the peak of the probability distribution remains well defined at all temperatures
after a Thouless cycle. It moves by an integer number of unit cells, determined by the finite-temperature topological invariant of the RMM. 
Because the initial state is however not an exact eigenstate, there are non-adiabatic correction during a pump cycle which result into a homogeneous offset in the particle distribution. Subtracting this offset and renormalizing the probability distribution leads to an exactly quantized shift of the COM.
Thus the discussed transfer scheme provides a tool to directly observe topological invariants of finite-temperature states, such as the ensemble geometric phase in non-interacting \cite{Linzner-PRB-2016,Bardyn-PRX-2018} systems.

%%%%%%%%%%%%%%%%%%%%%%%%%%%%%%%%%%%%%%%%%

%%%%%%%%%%%%%%%%%%%%%%%%%%%%%%%%%%%%%%%%%%%

%%%%%%%%%%%%%%%%%%%%%%%%%%%%%%%%%%
\subsection*{acknowledgement}
Financial support from the DFG through SFB TR 185, project number 277625399  is gratefully acknowledged.

%%%%%%%%%%%%%%%%%%%%%%%%%%%%%%%%%%%%%%%%%%%%%%%%%%%%%%%%%%%%%%%%%

\end{document}